\documentstyle{article}

\input psfig

\def\etal{{\it et\thinspace al.\thinspace}}

\def\lya{Lyman-$\alpha$\ }

\def\mincir{\ \raise -2.truept\hbox{\rlap{\hbox{$\sim$}}\raise5.truept
\hbox{$<$}\ }}
\def\magcir{\ \raise -2.truept\hbox{\rlap{\hbox{$\sim$}}\raise5.truept
 	\hbox{$>$}\ }}	
\def\hmpc{\,h^{-1}\,{\rm Mpc}}

\begin{document}
\large
\title{Probing the Large Scale Structure with QSOs and radio-galaxies
}
\author{Stefano Cristiani$^{1,2}$\\
{\it $^1$ Universit\`a di Padova, Vicolo dell'Osservatorio 5, I--35122 Padova}\\ 
{\it $^2$ESO, K. Schwarzschild-Stra{\ss}e 2, D--85740 Garching}}
\maketitle
\medskip
\centerline{Invited Review at the MPA/ESO Cosmology Conference} 
\centerline{"Evolution of
Large-Scale Structure: From Recombination to Garching"}
\centerline{Garching, Germany, 2-7 August, 1998.}
\medskip 

\begin{abstract}
{\large QSOs and radio-galaxies, together with the CMB,
``normal'' galaxies and clusters, 
represent the main source of
information about the origin and evolution of the Large Scale Structure. 
They can be
used either directly, as tracers of the density peaks, or as cosmic
lighthouses, probing the Universe along the line of sight with their
conspicuous flow of photons. Observations of the QSO and radio-galaxy 
clustering and
its evolution interweave information about their nature and the
cosmology that will hopefully be disentangled with the MAP
and PLANCK missions.
The amplitude of the QSO clustering has been measured to be $r_o \simeq
6 \hmpc$ at $z\sim 2$ and appears to increase with redshift. 
These properties are consistent with a
scenario in which the quasars, with a lifetime of $t \sim
10^8 yr$, sparsely sample halos of mass greater than $M_{min} \sim
10^{12}-10^{13}~h^{-1}$ M$_{\odot}$. 
Radio-galaxies are found to cluster on similar scales. The
study of QSO and radio-galaxy environments is in qualitative agreement
with their clustering properties.
Absorption-line systems observed in high-resolution spectra of QSOs show
structure, including voids/underdensities and overdensities, from the
smallest scales to tens of Mpc.
QSO pairs, groups and lenses allow us to get a 3-D glimpse of the absorbers,
providing invaluable information about their size, ionization, mass.
Great progress in this field is expected from the new instrumentation
coming into operation at the 8-m class telescopes.
Much work remains to be done also at low redshift to clarify whether  
the \lya absorption systems
have a direct physical connection with galaxies or a significant
fraction of them is caused by a truly intergalactic medium.
}
\end{abstract}

\markboth{S.Cristiani}{Probing the LSS with QSOs and radio-galaxies}

\section{Clustering of QSOs}

\subsection{Small scales}
The first indications of QSO clustering came from the
observations of possible physical pairs of quasars 
around 1980 (Hazard \etal 1979, Oort \etal 1981, Margon \etal 1981,
Webster 1982).
Although the individual cases were of tantalizing interest,
it was difficult to gain an appreciation of their
true significance from a-posteriori statistics. 

Systematic searches began with the pioneering work of Osmer (1981) and
the first detection was obtained by Shaver (1984), using a large,
inhomogeneous QSO catalog. The surveys carried out to investigate the
QSO Luminosity Function (LF) before long provided
statistically-well-defined samples 
with sufficient surface density and size for a study of the clustering
on scales of a few tens of Mpc (Shanks \etal 1983, 1987).

The two-point correlation function (TPCF) and its
Fourier transform, the power-spectrum, have been
applied as a standard approach 
to investigate the QSO clustering, but other techniques have also been
explored, such as counts in cells (Andreani \etal 1994, Carrera \etal
1998), the minimal spanning tree (Graham
\etal 1995), fractal analysis (Andreani \etal  1991) and the
friend-of-friend algorithm (Komberg \etal  1996).

Notwithstanding all the caveats due to their exotic nature, QSOs display a
number of appealing properties when compared to galaxies as
cosmological probes of the intermediate-linear regime of clustering,
on distances $\magcir 20$ Mpc: they show a rather flat redshift
distribution, make it possible to
define samples which are in practice locally volume limited, their
point-like images are less prone to the surface-brightness biases
typical of galaxies and they sparse-sample the environment.

In recent times complete samples totaling about 2000 QSOs have been
assembled, providing a $\simeq 5 \sigma$ detection of the QSO
clustering on scales of the order of $6 h^{-1}$ comoving Mpc (Andreani
\& Cristiani 1992, Mo \& Fang 1993, Croom \& Shanks 1996), an
amplitude which appears to be consistent with or slightly larger than
what is observed for present-day galaxies and definitely lower than
the clustering of clusters.

The evolution of the QSO clustering with redshift is still a matter of
debate.  An amplitude constant in comoving coordinates or marginally
decreasing with increasing redshift has been the standard claim till
recently.  However, a number of indications currently point the
opposite direction.  The data of the Palomar Transit Grism Survey
(Kundi\'{c} 1997, Stephens \etal 1997) make it possible to measure the
amplitude of the TPCF at redshifts higher than 2.7 and the result,
$r_o = (18\pm8) h^{-1}$ Mpc, is suggestively higher than what is observed
at lower redshifts.

La Franca \etal (1998) have observed a sample of 388 QSOs with
$0.3<z<2.2$ and $B_j \leq 20.5$ over a {\it connected} area of 25
square degrees in the South Galactic Pole (SGP).  The TPCF analysis
gives an amplitude $r_o = (6.2\pm1.6) ~h^{-1}$ Mpc, in agreement with
previous results. But when the evolution of the clustering with
redshift is analyzed, evidence is found for an {\it increase} of the
clustering with increasing z, although only 
at a $2 \sigma$ level.

\begin{figure}[t]
\centering\mbox{\psfig{figure=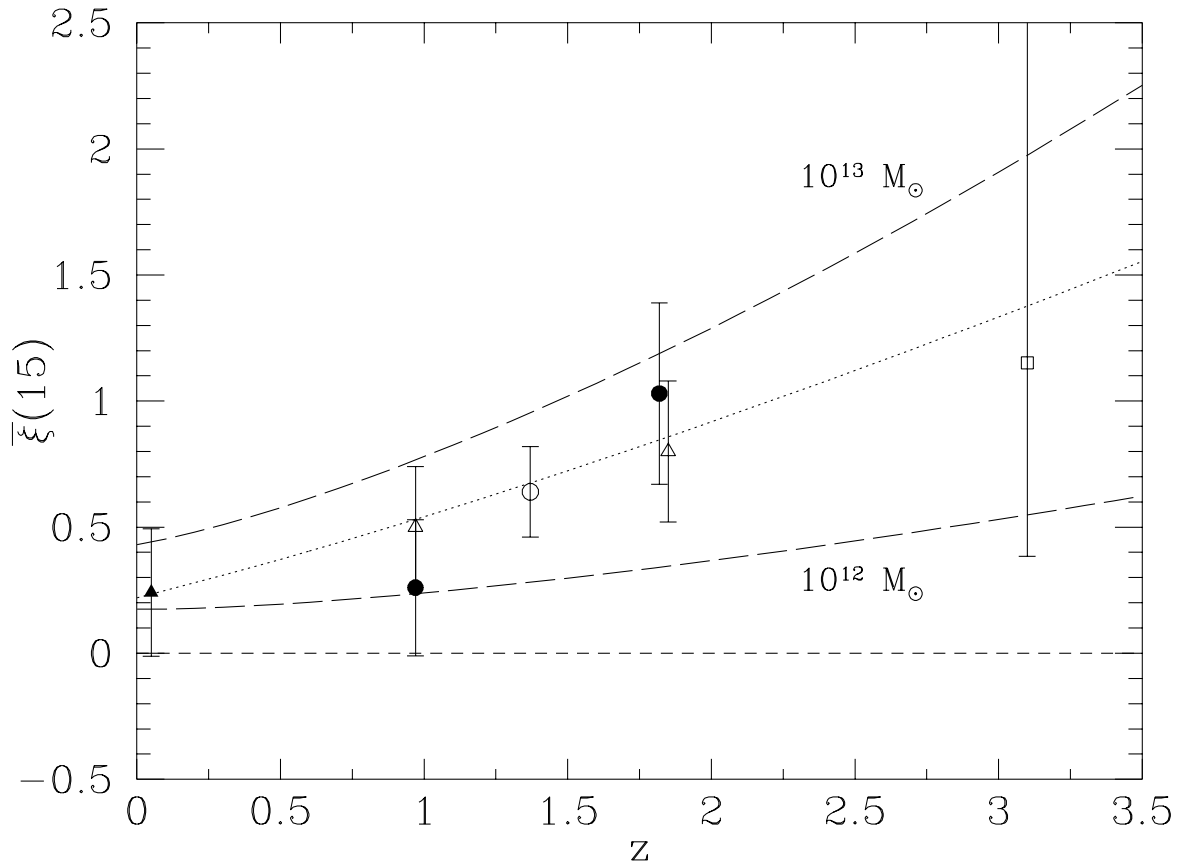,height=12cm}}
\caption[]{
The amplitude of the $\bar\xi(15~h^{-1}$ Mpc) as a function of z.
Filled circles: the low- and high-$z$ SGP subsamples (La Franca \etal
1998); open circle: the SGP sample plus other optical samples
available in the literature (see La Franca \etal 1998 for a detailed
list); open triangles: same as open circle but divided into two redshift
slices; filled triangle: low-$z$ AGNs from Boyle and Mo (1993) and
Georgantopoulos and Shanks (1994); open square: the high-$z$ sample
of Kundi\'{c} 1997. The dotted line is the $\epsilon= -2.5$
clustering evolution fitted to the open triangles and the filled
triangle data.  The dashed lines are the $10^{12}$ and $10^{13}$
$M_{\odot}$ $h^{-1}$ minimum halo masses clustering evolution
according to the transient model of Matarrese \etal  (1997).
} \label{QSOclustev_fig}
\end{figure}

Boyle \& Mo (1993) measured the clustering of low-z QSOs in the EMSS
and Georgantopoulos \& Shanks (1994) used the IRAS point source
catalog to measure the clustering of Seyferts. Altogether a low value
of the average TPCF within 15 Mpc and z=0.05 is obtained, $\bar\xi = 0.25 \pm
0.24$.

On the other hand, Carrera \etal (1998), on the basis of 235 X-ray
selected AGNs from the RIXOS and DRS surveys, do not support an
increase of the clustering amplitude with redshift.  It is also worth
noting that the clustering measured by Carrera \etal (1998) for these
X-ray selected objects, $1.9 \mincir r_0 \mincir 4.8\hmpc$ (with the
non-crucial assumption of stable evolution), is significantly smaller
than what is observed for optically-UV-selected QSOs.  This opens the
interesting possibility of having a contribution of AGNs to the XRB
larger than the standardly quoted 50\%, without violating the limits
imposed by the angular auto-correlation-function (Carrera \& Barcons
1992, Carrera \etal 1993, Georgantopoulos \etal 1993, Danese \etal
1993, Chen \etal 1994, So\l{}tan \& Hasinger 1994).

The customary appeal ``more data are needed'' arises naturally in this
context: hidden correlations, for example with the absolute
luminosity, as observed for galaxies (Park \etal 1994), although
undetected or undetectable in the present data, might confuse our
perspective on the issue of clustering evolution.  
Hopefully we will not wait long, thanks to
the 2dF QSO redshift survey (cf. Croom \etal, these proceedings).

At the moment, if we limit the analysis to the optical data and
parameterize the evolving correlation function in a standard way:
$$\xi(r,z) = ({r/{r_0}})^{-\gamma}(1+z)^{-(3-\gamma+\epsilon)}$$ where
$\epsilon$ is an arbitrary (and not very physical) fitting parameter,
we obtain $\epsilon = - 2.5\pm 1.0$, which appears inconsistent with
the $\epsilon \simeq 0.8$ observed for faint galaxies at lower
redshifts (Le F\`{e}vre \etal 1996, Carlberg \etal 1997, Villumsen
{\etal} 1997, Postman {\etal} 1998).  Great care should be exercised
however when carrying out this comparison. Quasars could be more
significantly related to the population of Lyman-break galaxies, which
show substantial clustering at $z \simeq 3.1$ (Steidel \etal 1998) and
can be interpreted as progenitors of massive galaxies at the present
epoch or precursors of present day cluster galaxies (Governato \etal
1998).

As mentioned, the standard parameterization of the clustering may
not be a sensible description.  The observed clustering is the result
of an interplay between the clustering of mass, decreasing with
increasing redshift, and the bias, increasing with redshift, convolved
with the redshift distribution of the objects (Moscardini \etal 1998,
Bagla 1998) and obviously is critically linked to the physics of
QSO formation and evolution.  Let us consider here as an archetypical
scenario the one outlined by Cavaliere \etal (1997), modeling the
{\it rise and fall} of the QSO LF as the effect of two components: the
newly formed black holes, which are dominant at $z>3$ and the
reactivated black holes,
which dominate at $z<3$.  The reactivation is triggered by
interactions taking place preferentially in groups of galaxies of
typical halo mass $\sim 5 \cdot 10^{12}$ M$_\odot$.

In this way the clustering properties of QSOs are related to those of
transient, short-lived, objects (Matarrese \etal 1997). 
This is markedly different from the situation with respect to
galaxies, which in turn can be associated with a merging or
object-conserving paradigm of long-lived objects.  At high redshifts
QSOs correspond to larger (rarer) mass over-densities collapsing early
and cluster very strongly.  Then the clustering amplitude decreases
till the mass scale typical of a QSO reaches the value of the average
collapsing peaks, after which clustering may grow again. 

If we think of QSOs as objects sparsely sampling halos with $M >
M_{\rm min}$, an $M_{\rm min}= 10^{12} - 10^{13}~
h^{-1} M_{\odot}$ would provide the desired amount {\it and} evolution of
the clustering.
\subsection{Quasar-galaxy clustering}
The ideas presented in the previous section are corroborated by the
study of QSO environments.  Over the years, starting with the
pioneering observations of Stockton (1978, 1982), evidence has been
accumulating that QSOs are associated with enhancements in the galaxy
distribution.  At low redshifts ($z \mincir 0.5$) quasars reside in
small to moderate groups of galaxies, rather than in rich clusters
(Yee \& Green 1984, Haymann 1990, Smith \& Heckman 1990).  The
QSO/galaxy correlation function has been measured to be $3.8 \pm 0.8$
times stronger than the galaxy/galaxy correlation function (Fisher et
al. 1996).  At higher redshifts, there is a marked difference in the
environments of radio-loud and radio-quiet QSOs (Yee \& Green 1987,
Yee \& Ellingson 1993). Radio-loud QSOs are often found in association
with rich clusters (Abell richness $\geq 1$), at least on scales
larger than $\sim 0.5 \hmpc$ (Hall \& Green 1998), while radio-quiet
quasars appear to remain in smaller groups.

An indirect confirmation comes from the observations of close
line-of-sight (LOS) QSO pairs. In their absorption spectra lines are often
seen in the spectrum of the more distant QSO at a redshift corresponding to
that of the foreground QSO - the so-called {\it associated absorption} -
showing that QSOs are located in higher-than-average galaxy density
regions (Shaver \& Robertson 1983, Phillips 1986, Cristiani \& Shaver
1988).

\subsection{Large Scales}
The study of the QSO clustering on scales larger than a few tens of
Mpc is a much less beaten path. On the one hand the familiar TPCF is
known to be inefficient for detecting structures on scales comparable
with the dimension of the sample (e.g. Pando \& Fang 1996) and other
approaches, such as the wavelet analysis or counts-in-cells, may
represent better tools.
On the other hand observational biases may be
extremely dangerous in this situation: for example modulations of the
sensitivity on scales of the order of the size of a Schmidt plate may
introduce spurious power on scales of the order of 100 Mpc.  In the
present absence of proper systematic surveys on these large scales one
should therefore take with the classical grain of salt the several
examples of QSO super-structures, derived mostly from inhomogeneous
catalogues, being aware of the pitfalls of a-posteriori
statistics. For the large scales we are in many respects in a
situation similar to that of the early eighties for the small
scales.

Nonetheless there are at least two known cases which appear as
outstanding examples of large quasar groups:
\begin{enumerate}
\item {a group of 23 quasars within $\sim 60 \hmpc $ at $z\sim 1.1$, 
found by Crampton, Cowley \& Hartwick (1989)}
\item {a group of 14 quasars within $\sim 100 \hmpc $ at $z\sim 1.9$,
found by Komberg {\etal} in the SGP.
In the SGP, thanks to the large number of deep surveys carried out by
various authors, at least a couple of other possible super-structures, although
of less compelling evidence, have been identified (Clowes and
Campusano 1991, Graham \etal 1995).}
\end{enumerate}

Such large groups appear as possible progenitors of
the present-epoch great attractors (i.e. regions in the local Universe
with over-densities $ \magcir 10$ on scales $\simeq 100 \hmpc$):
and their spatial comoving density can be estimated to be comparable
to local super-clusters. 
Pinpointing the QSO clustering on these large scales would immediately
provide another challenge to cosmological models, but for this 
a systematization of the field is needed.
Great advances are expected from the wide-field facilities which have come recently into operation:
SDSS, the 2dF QSO survey and
also from large-area follow-up spectroscopy of radio surveys like FIRST.

\section{Clustering of Radio-Galaxies}
Inspired by the paradigm unifying radio-loud QSOs and
FRII radio-sources it is interesting to check whether the
observations of radio-galaxies measure clustering properties
consistent with those of QSOs.

Historically, the long debate on the spatial distribution of radio
galaxies led to the widely-accepted view that radio sources were
isotropically distributed, at least on large scales (Webster
1976). \\ 
The problem with radio surveys is that in general they lack
the redshift information and the clustering analysis has to resort to
the angular correlation function.  
The large range in intrinsic luminosity of radio
sources leads then to a wide convolution that dilutes dramatically any
clustering signal.  In fact, when the studies of the radio sources
have been limited to nearby objects as Seldner \& Peebles (1981) did
for 4C sources at $z<0.1$ or Shaver \& Pierre (1989) for Molonglo
sources at $z \sim 0.02$, structures were found, including the
super-galactic plane.  Peacock \& Nicholson in 1991 used an all-sky
sample of radio-galaxies at $z<0.1$ to measure their 3D TPCF,
$r_0\simeq 11h^{-1}$Mpc.
In recent times Kooiman \etal (1995) and
Loan, Wall \& Lahav (1997) analyzed the 4.85 Ghz Green Bank and the
Parkes-MIT-NRAO surveys with the TPCF, showing that radio objects
appear to be more strongly clustered than local optically-selected
galaxies.\\
Magliocchetti (this conference) reports on the studies carried
out on the FIRST database (Magliocchetti \etal  1998a,b, Cress et al
1996,1998), which provide indeed an estimate of the correlation
amplitude, $r_0\sim 9-11h^{-1}$Mpc, for objects typically at $z \sim 1$,
consistent with the QSO result.
To gain more robust insight it is however mandatory
to obtain redshifts and radio morphologies at least for a
training subsample of these databases in order to remove the basic
statistical uncertainties concerning the redshift distribution and the
multi-component nature of many sources.

The environments of radio-galaxies at redshifts $z \sim 0.5 - 1$ also
appear to be remarkably similar to those of QSOs, that is {\it on
average} not richer than Abell class 0 clusters
(Roche \etal 1998), with a correlation
function significantly stronger ($r_0\sim 8h^{-1}$Mpc) and steeper
($\gamma \sim 2$) than what is observed for ``normal'' galaxies, but
consistent with the estimates for local early-type galaxies of Guzzo
\etal (1997).

\begin{figure}[t]
\centering\mbox{\psfig{figure=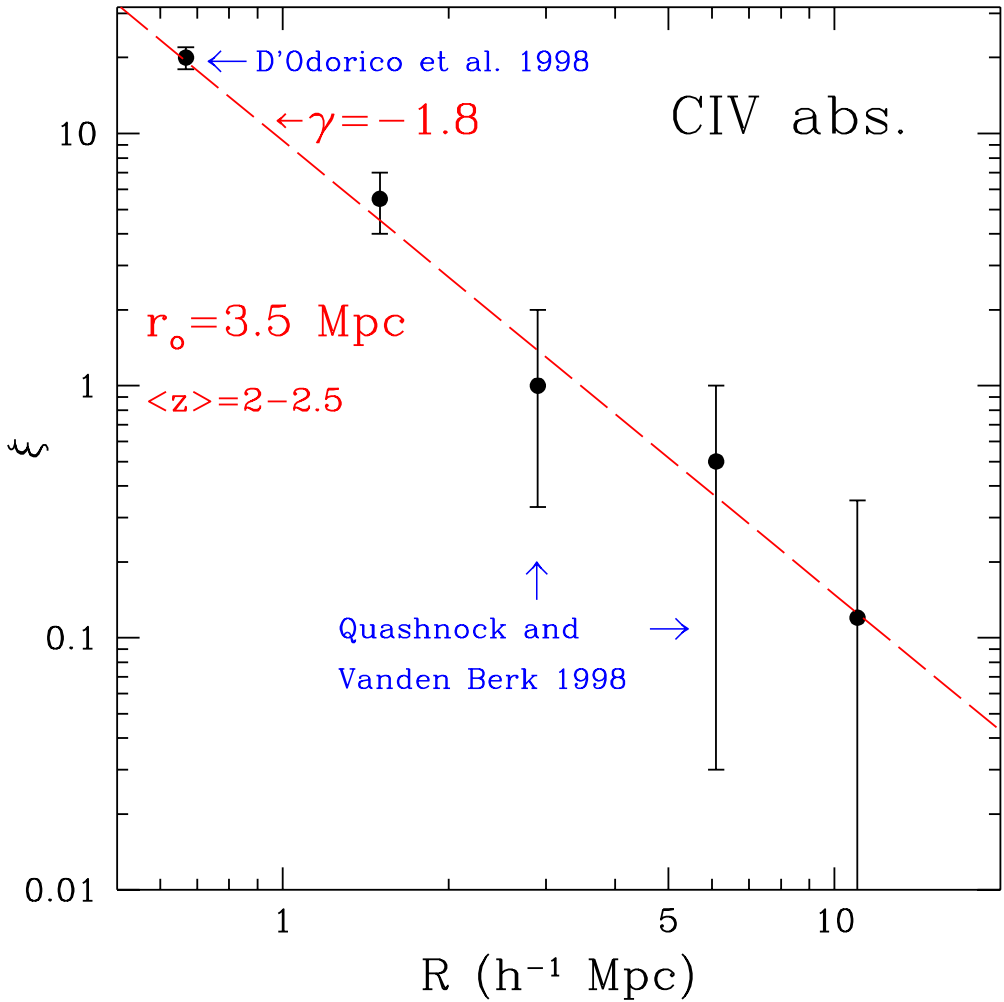,height=14cm}}
\caption[]{The two-point correlation
function of CIV systems as measured by Quashnock and Vanden Berk (1998)
and by D'Odorico \etal  (1998a)
} \label{figclustCIV}
\end{figure}

\section{Clustering of Absorbers}
A complementary view of the LSS is provided by the study of QSO
absorption spectra, which sensitively probe, in a luminosity unbiased way, 
the spatial distribution of baryons in the universe.

\subsection{Metal lines}
Historically, the so-called sharp metal-lines have been the first
class of absorbers recognized to cluster significantly.
Already in 1982 Young, Sargent \& Boksenberg indicated that the peak
observed in the TPCF of CIV doublets could be due to galaxy-galaxy
clustering at $z\sim 2$ rather than to multiple clouds in individual
galaxies. This was reiterated in 1988 by Sargent, Steidel \& Boksenberg 
on the basis of a sample of 55 QSOs observed at $100$ km/s
resolution, measuring a correlation function $\xi \sim 5-10$ for LOS
velocity differences $\Delta v = 200-600$ km/s.
In recent times Petitjean \& Bergeron (1994), Songaila \& Cowie (1996)
and D'Odorico \etal (1998a), using higher-resolution, smaller samples,
have measured a  $\xi \sim 20 $ at $\Delta v \simeq 120$ km/s.
Quashnock \& Vanden Berk (1998) have used a compilation of heavy-element QSO
absorbers to estimate the CIV TPCF from 200 up to about 3000 km/s.
The various authors unanimously remark that on small scales the
velocity structure must to a certain extent reflect relative motions
of clouds within individual halos, a circumstance confirmed by the
relative weakness of the correlation observed in the transverse
direction on the adjacent
lines-of-sight of a QSO triplet (Crotts \etal 1997). 
However, we can see from Fig. 2 that, taken at face value, the results by
D'Odorico \etal on small scales and by Quashnock \& Vanden Berk together
define with remarkable accuracy a CIV CF at $z\simeq 2-2.5$ of the
canonical power-law form, with an $r_o = 3.5$ Mpc and $\gamma \simeq
1.8$
. \\
An evolution of the CF as a function of $z$ is also detected and can be
described in the standard parameterization of \S 1 by an $\epsilon
= 2 \pm 1$. 
This evolutionary pattern is apparently followed also by the MgII
systems, which, as shown by Petitjean \& Bergeron (1990, 1994) and
Caulet (1989), in spite of the different ionization phase are 
similar to CIV systems in terms of cross sections.

There is also evidence for a dependence (increase) of the clustering
strength on the (increasing) column density of the systems (D'Odorico
\etal 1998a, Quashnock \& Vanden Berk 1998), which suggests, together
with the large correlation scale found, that the CIV absorbers are
biased tracers of relatively high density regions.
It also cautions us that the agreement with the standard $\gamma \simeq 1.8$
power-law behavior of the CF shown in Fig.2 may be
``too-good-to-be-true'', since the lines of the sample by D'Odorico et
al. have on average lower $N_{CIV}$ ($EW\sim 0.1$\AA) than the ones in
the Quashnock \& Vanden Berk sample 
($EW\sim 0.4$\AA). Again, plenty of data are needed to disentangle the
various dependences, including the evolutionary rates as a function of
the column density.

\subsection{Lyman Forest}

The search for clustering of the Lyman-$\alpha$ lines has produced
disparate results over the years.  Systematic studies of the redshift
distribution in the QSO Lyman-$\alpha$ forest began in
the early 80's with the work by Sargent \etal (1980) who
concluded that no structures could be identified.  Almost all the
subsequent results have failed to detect any significant correlation
on velocity scales $300 < \Delta v < 30000$ km s$^{-1}$
(Sargent \etal 1982, Webb \& Barcons 1991).
On smaller scales ($\Delta v =50-300$
km s$^{-1}$) there have been indications of weak clustering
(Webb 1987, Crotts 1989, Rauch \etal 1992, Chernomordik 1995,
Cristiani \etal 1995, Kim \etal 1997)
together with relevant non-detections (Pettini \etal 1990, 
Stangler \& Webb 1993, Kirkman \& Tytler 1997).
The absence of power in the TPCF
has been claimed as a striking characteristic of
the Lyman-$\alpha$ forest and has been used,
together with the observed low metallicity, as a basic argument to
develop a scenario of the Lyman-$\alpha$ absorbers as a totally
distinct population with respect to metal systems and therefore
galaxies.

\begin{figure}[t]
\centering\mbox{\psfig{figure=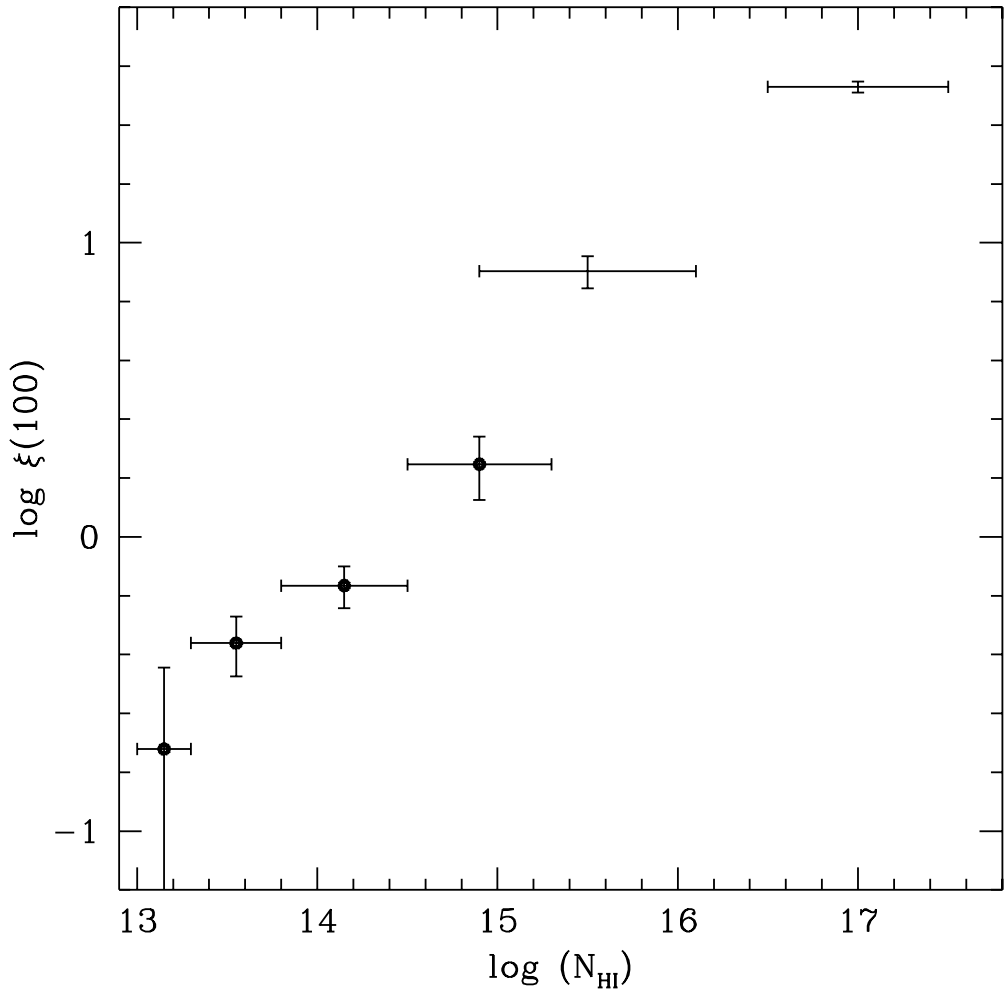,height=14cm}}
\caption[]{ Variation of the amplitude of the TPCF
as a function of the column density threshold for the sample
of the Lyman-$\alpha$ lines (filled circles).  The two points in the
upper-right area of the picture show the correlation of the CIV metal
systems (see text)} \label{figclustew}
\end{figure}

In recent times a large database of high-resolution ($\sim 10$ km/s),
high-S/N spectra of the Lyman forest has become available, allowing a
detailed investigation, especially at high
redshift, where the density of lines provides particular sensitivity.
Cristiani \etal (1997) have used more than 1600 \lya lines
to detect a weak but significant signal, with $\xi\simeq 0.2$ on
scales of 100 km s$^{-1}$ at a $4.6 \sigma$ level.

Exploring the variations of the clustering as a function of the column
density no evidence for clustering is found for lines with $\log
N_{HI}\le 13.6$.  On the contrary, for lines with $\log N_{HI}\magcir
14$, the correlation function at $\Delta v = 100$ km s$^{-1}$ shows a
remarkable increase in amplitude ($\xi\simeq 0.7$) and significance
($>7 \sigma$).  No relevant feature other than the peak at small
velocity separations is observed. In particular previous claims for
anti-clustering on scales $\sim 600 - 1000$ km s$^{-1}$ are not
confirmed.  Fig.~\ref{figclustew} shows more in detail the variation
of the amplitude of the TPCF as a function of the
column density.  A trend of increasing amplitude with increasing
column density is apparent all the way up to the two points in the
upper right that represent the clustering of metal systems.

The tricky part in the otherwise trivial computation of the TPCF in the
velocity space
$$ \xi(\Delta v) = {N_{obs}(\Delta v) \over N_{exp}(\Delta v)} -1$$
is the correct estimation of $N_{exp}$, the number of pairs
expected in a given velocity bin from a random distribution in
redshift.
The results are not sensitive to the particular cosmological evolution
($\propto (1+z)^{\gamma}$) adopted, but the effects of 
the intrinsic blending, line blanketing of weak lines due to strong
complexes and finite spectral resolution have to be carefully reproduced.

Fern\'andez-Soto \etal (1996) have investigated the clustering
properties of the Lyman-$\alpha$ clouds on the basis of the
corresponding CIV absorptions.  They measure in this way a strong
clustering for lines with $N_{HI} > 3 \cdot 10^{14}$ cm$^{-2}$,
suggesting that CIV may resolve better the small-scale velocity
structure. A caveat however has to be filed, about
the different behavior of high and low-ionization species and 
in general of \lya and metal systems
(see also below the observations of QSO gravitational lenses, \S 3.3).

Of course, rather than just correlating lines regardless of their
properties, increasingly more refined approaches are possible:
from the correlation of EW or $N_{HI}$ weighting (Webb \&
Barcons 1991, Zuo 1992), to the correlation of optical depths or pixel
intensities (Press \etal 1993, Cen \etal 1998), to the full recovery
of the density field (Haehnelt \& Nusser, this conference) and much is
being learnt from simulations (Weinberg, this conference).

The study of the evolution of the TPCF with the redshift in the optical
range ($1.7 < z < 4$) suggests that for Lyman-$\alpha$ lines with
column densities 
$\log(N_{HI}) > 13.8$ the amplitude of the correlation at $100$ km
s$^{-1}$ decreases with increasing redshift (Cristiani \etal 1997).
Unfortunately, HST data are still at too low-resolution or are too
scanty (Bahcall \etal 1996, Brandt \etal 1995) to allow a meaningful
comparison.  Nonetheless, the result by Ulmer (1996), who measured
with FOS data a $\xi = 1.8^{+1.6}_{-1.2}$, on scales of $200-500$ km/s
at $0 < z < 1.3$, suggests that the trend persists at lower redshifts.

If we add to this observation the following pieces of evidence:
\begin{enumerate}
\item At low redshift a significant fraction of the Lyman-$\alpha$
lines has been observed to be associated with luminous galaxies and
the local large-scale structure (Le Brun \etal 1996, Chen \etal 1998,
\S 4);
\item Metallicities of the order $10^{-2}$, i.e.  similar to the ones
derived for the heavy-element absorptions originated in galactic
halos, are observed for Lyman-$\alpha$ clouds with $\log N_{HI} > 14$
(Cowie \etal 1995, Songaila \& Cowie 1996);
\item The volume density and cosmological evolution of the same $\log
N_{HI} > 14$ clouds are similar to those of the damped systems
(Giallongo \etal 1996);
\end{enumerate}
all together this suggests a physical association between the
Lyman-$\alpha$ clouds with $\log N_{HI} > 14$ and the halos of
protogalactic systems.

The typical column density above which the clustering is
observed ($\log (N_{HI}) \sim 14$) corresponds to the position of the
break in their column density distribution (Petitjean \etal 1993,
Giallongo \etal 1996) and we should not be surprised that below that
no clustering is observed since the weaker lines are recognized to
probe under-dense regions of the universe.

\subsection{Large-scale Clustering}

On large scales observations show the existence of significant
over- and under-densities of lines on scales of a few tens of Mpc.  A
typical case is the occurrence of the so-called voids for which a few
interesting claims exist: Crotts (1987, 1989) was the first to suggest
the presence of a $\sim 40 \hmpc$ gap of absorption lines in the
Lyman forest of the QSO 0420-388, Dobrzycki \& Bechtold (1991a, 1991b)
found a void of $32 \hmpc$ in Q0302-003 and Cristiani \etal (1995,
1997) two of about $20 \hmpc$ in Q0055-269.  The systematical analysis
of the same sample of \lya lines described in the previous subsection
corroborates the idea that voids of a few tens of Mpc occur
occasionally in QSO absorption spectra: the filling factor is low, a few
percent of the available LOS path-length.  Regions corresponding to
the voids are not completely devoid of lines: weak absorptions are
observed within the voids. This agrees with low-redshift observations
(Shull \etal 1996) showing that in the local Universe voids
are not entirely devoid of matter.

Over densities on scales again of a few tens of Mpc (never more than that)
have been found at $z=1.5-4$, roughly with the same frequency observed
for under-densities.

The most remarkable ones remain Q0237-233 (Sargent \etal 1988, 
Heisler, Hogan \& White 1989) 
and the pair 1037/1038-27 (Jakobsen \etal 1986, 1988, Cristiani
\etal 1987, Sargent \& Steidel 1987, Lespine \& Petitjean 1997), 
which display clumps of
metal-lines on scales of the order of 10000 km/s.  Francis \&
Hewett found an interesting coincidence (within 300 km/s) of two pairs
damped systems at z=2.38 and 2.85 in the spectra of two QSOs whose LOS
are separated by $6 \hmpc$.

\begin{figure}[t]
\centering\mbox{\psfig{figure=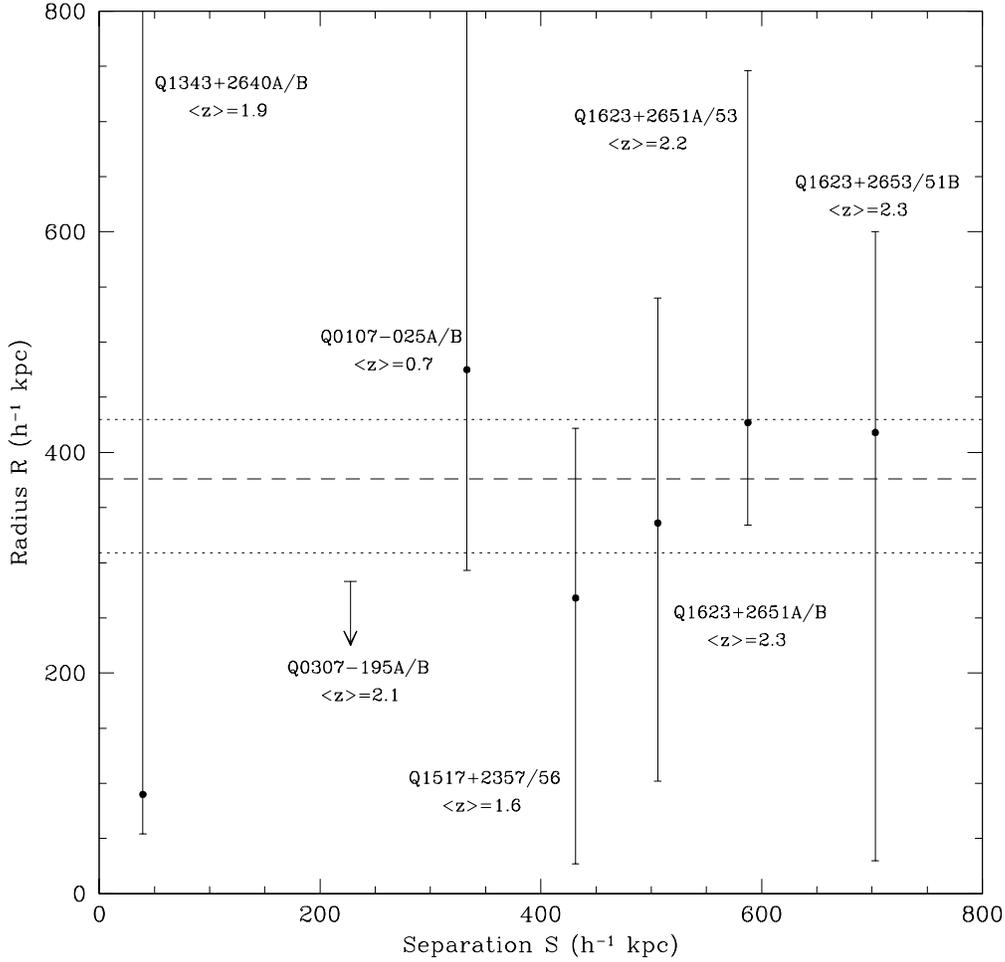,height=14cm}}
\caption[]{ 
Inferred size of \lya absorbers -
from a model assuming single-radius $R$, unclustered 
spherical clouds - as a function of QSO pair sightline separation $S$. 
The dashed line shows the most probable value for $R$ and the dotted lines 
the 95 \% confidence interval 
(from D'Odorico \etal 1998b)
} \label{pairs}
\end{figure}

Great care should be exercised before interpreting
these structures in terms of gravitational instabilities since other
phenomena, also of great interest, may be acting: proximity
effects, i.e. excess ionization caused by a nearby QSO, and any
ionization inhomogeneity in the UV background which becomes more and
more important with increasing redshift (Haardt and Madau 1996,
Reimers \etal 1998).

\subsection{Cross-correlation of different lines of sight:
QSO pairs and groups.}

An unfortunate shortcoming of clustering studies based on absorption
lines is the 2-D nature of the information, typically confined to the
analysis along the LOS.  Pairs and groups of QSOs offer an invaluable
insight in this respect and the advent of 8-m class telescopes will
certainly boost the field, allowing a systematic detailed 3-D mapping
of absorbers and galaxies at the same time up to the highest
redshifts.

Historically, quasar pairs have provided information about
{\it a)} the size of the \lya absorbers, which turned out to be much
larger than the few tens of kpc scale thought a decade ago,
{\it b)} the existence of
superclusters at high redshift (see \S 3.3),
{\it c)} the association of QSOs with moderate overdensities (\S 1.3).
 
Where do we stand now?
On small scales ($\sim 1$ kpc) gravitational lenses have shown that
the forest of \lya lines in the spectra of the two images of a lens can be
amazingly similar with differences less than a few percent on scales of
a few hundred pc (Rauch, this conference). This is very important
because it tells us that in the case of \lya - not metal lines - the
medium is quite homogeneous on such scales and therefore an empirical
limit is provided for the critical issue of the 
resolution in hydrodynamical simulations (cf. Theuns, this conference).

At the intermediate ($O({\rm arcmin})$) scales progress has been slow
in the last decade, due to the scarcity of suitably bright pairs and
groups for 4m-class telescopes.  D'Odorico \etal (1998b) have recently
re-observed the QSO pair 0307-195 and re-discussed this and the other
7 pairs in the literature with data of comparable quality.  A typical
size of the absorbers - with the assumption of spherical geometry - of
$360 h^{-1}$ kpc is inferred (fig.4).  Besides, remarkably quiet
velocity fields are observed with differences between
coincident lines within 100 km/s (Dinshaw \etal 1994).  The present
data are not sufficient to support a correlation of the typical
inferred size with the proper separation or with the redshift of the
pairs, which was claimed in the past.  

It may be argued then whether such dimensions can really be
considered cloud radii or are rather correlation lengths, a
distinction which tends to be blurred in view of the present
hydrodynamic simulations.  It is also debatable whether flattened
disk-like geometries tend to be favored (Dinshaw \etal 1997), or
spherical clouds with an $r^{-4}$ column density profile and a
power-law distribution of radii can give a satisfactory representation
of the observations, too.
Certainly the empirical evidence points to large ionization fractions
and consequently large masses. On the basis of the inferred sizes and
the UV background
measured at $z=2-4$ by Giallongo \etal (1996) the optically thin \lya
absorbers probed in QSO pairs are expected to have a neutral fraction
$\mincir 10^{-5}$ and their contribution to the mean intergalactic
density is close to conflict with the baryon limit from
nucleosynthesis.

On the large scales, after the studies of TOL1037/1038, there has been
disappointingly little progress. The most
remarkable recent result has been produced by Williger \etal (1996) who
observed metal systems in
25 LOS spanning $1.5 < z < 2.8$, spread over about 1 sq. degree
in the South Galactic Pole.
In spite of the praiseworthy effort, the final sample of CIV systems 
collected by Williger \etal makes it possible to compute the TPCF by
cross-correlating a maximum of 22 absorbers and is just sufficient to
reject the null hypothesis of a random distribution.
The reason is again the same: the surface density of QSOs accessible
to 4-m telescopes for this type of work is simply too small.

Future prospects, however, look really exciting.  Starting at the end of
1999, the echelle spectrograph UVES will be available on the UT2 of
VLT, extending, with respect to HIRES at Keck, the possibility of
high-resolution ($ \Re \sim 50000$) 
observations down to at least $V\sim 20$ and to a
shorter wavelength range (i.e. to more numerous, lower-redshift
QSOs).  Then (by 2001) the FLAMES facility (see {\tt
http://http.hq.eso.org/instrumen\\
ts/flames/}) will allow observation
with UVES of up to 8 QSOs at the same time in a field of 25 arcmin
diameter and/or 130 targets in the same area with an intermediate/high
resolution spectrograph, GIRAFFE.  This opens up enormous possibilities:
for example the Alcock-Paczinsky test applied to QSO groups to
investigate the cosmological geometry (Hui, this conference) would
require few observing nights.  This test, based on the comparison
between the clustering properties of the (\lya) absorbers, namely the
TPCF, observed along the LOS and the corresponding estimate in the
transverse direction, is especially sensitive to $\Omega_{\Lambda}$
and should be able to discriminate between an ($\Omega_m = 0.3,~
\Lambda=0$) and an ($\Omega_m = 0.7,~ \Lambda=0.3$) universe at a $4
\sigma$ level by observing $\sim 25$ QSO pairs (Hui
\etal 1998).

\section{Low-z Lyman absorbers and LSS}

At lower redshift, after the pioneering work on the identification of
galaxies giving origin to metal absorbers (Bergeron \& Boiss\'e 1991,
Bergeron \etal 1992), the
more recent activity on \lya systems, triggered by the HST
observations, is fueling a lively debate.  
A number of authors (Morris
\etal 1993, Lanzetta \etal 1995, Le Brun, Bergeron \& Boiss\'e 1996,
van Gorkom \etal 1996, Chen \etal 1998, Weymann et
al. 1998) have argued whether the low-redshift \lya absorption systems
have a direct physical connection with galaxies or a significant
fraction of them is caused by a truly intergalactic medium.

A central point in this debate is the anti-correlation between the
equivalent width and the impact parameter of the LOS from the center of
the galaxies measured by some authors (Lanzetta \etal 1995, Cen \etal
1998) and not fully confirmed by others (Le Brun \etal 1996, Bowen 
\etal 1996).

At the moment it appears possible to reconcile the various
observations {\it if} the presence of well defined galactic envelopes
is invoked only for the relatively strong absorption lines.  The
covering factor around each galaxy for this class of absorbers is high
at impact parameters $<200-300 h^{-1}$ kpc, dropping rapidly at larger
separations.  The weaker lines, which also display a significantly
slower redshift evolution (Weymann \etal 1998), occur then in
filaments up to several Mpc away from the nearest galaxy.

The zoo of 
interpretations for the low-z Ly$\alpha$ systems includes:
tidal tails (Morris \& Van den Berg 1994), galactic winds (Wang 1995),
outer parts of disk and irregular galaxies (Maloney 1992), clouds that
are pressure confined by a hotter gas in extended galactic halos (Mo
1994), extended low redshift disk galaxies which, in a short burst of
star formation, blow out their denser centers and disappear from view,
while their gas cross-section remains (the ``Cheshire Cat model'':
Salpeter 1993), galaxy clusters (Lanzetta et al 1996).
Possibly, more than one population will have to be invoked
in order to fully account for all the observations.


\end{document}